\documentclass[twocolumn,floats,showpacs,pra]{revtex4}
\usepackage[above]{placeins}
\usepackage{amsmath}
\usepackage{graphicx}

\begin{document}

\title{What will be the maximum $T_{c}$ in the iron-based superconductors? }
\author{Xiuqing Huang$^{1,2}$}
\email{xqhuang@nju.edu.cn}
\affiliation{$^1$Department of Physics and National Laboratory of Solid State
Microstructure, Nanjing University, Nanjing 210093, China \\
$^{2}$ Department of Telecommunications Engineering ICE, PLAUST, Nanjing
210016, China}
\date{\today}

\begin{abstract}
Using the newly developed real space vortex-lattice based theory of
superconductivity, we study the maximum superconducting transition
temperature ($T_{c}^{\max }$) in the iron-based superconductors. We argue
that the $c$-axis lattice constant plays a key role in raising the $%
T_{c}^{\max }$ of the superconductors. It is found that all the reported
FeAs superconductors can be divided into two basic classes ($c/a\approx 3$
and $c/a\approx 5/2$) depending on the lattice constants, where $a$ is the
Fe-Fe distance in the $xy$-plane and $c$ is the Fe-Fe layer distance along
the $z$-axis. Our results suggest that the former class has a maximum $%
T_{c}^{\max }<60$ K, while the latter class has a lower $T_{c}^{\max }\leq
40 $ K. Our investigations further indicate that, in order to enhance $%
T_{c}^{\max }$ in this family of compounds, new class of superconductors
with a larger ratio of $c/a$ should be synthesized. It is likely that their $%
T_{c}^{\max }$ values could be raised into the liquid nitrogen range ($77$
K) and $100$ K, supposing the new analogues with $c/a\approx 5$
(approximately $c>13$ $\mathring{A}$, if $a=2.750$ $\mathring{A}$) and $%
c/a\approx 11$ ($c>31$ $\mathring{A}$) can be experimentally achieved,
respectively. For the new FeSe series, our mechanism predicts that their $%
T_{c}^{\max }$ is impossible to exceed $30$ K due to a relatively shorter $c$%
-axis lattice constant ($c/a\approx 2$). Finally, based on the new
experimental results (arXiv:0811.0094 and arXiv:0811.2205), the possible
ways to raise the $T_{c}$ of the iron-based superconductors into 70 K are
also suggested.
\end{abstract}

\maketitle

\section{Introduction}

\begin{figure}[tp]
\begin{center}
\resizebox{0.97\columnwidth}{!}{
\includegraphics{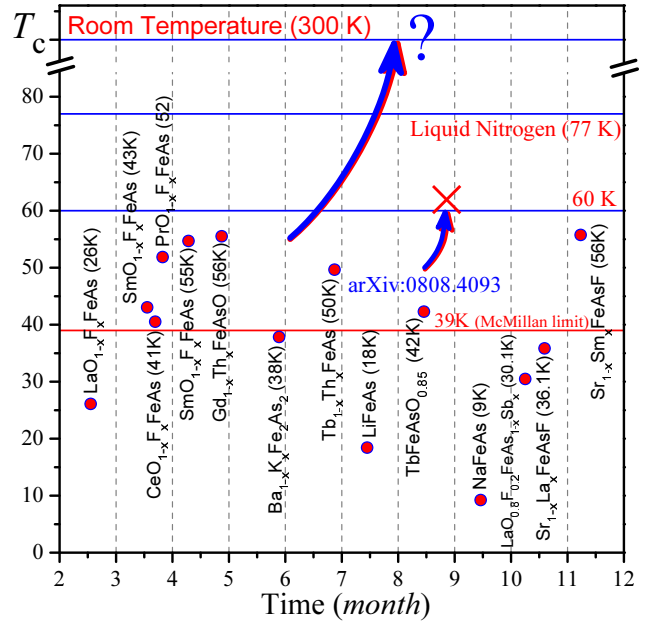}}
\end{center}
\caption{The schematic plot of the discovery of iron-based superconductors
in 2008. We argue that all the reported FeAs superconductors can be divided
into two basic classes (large $c$-axis and short $c$-axis), and the
corresponding $T_{c}$ values are very difficult to exceed 60 K. }
\label{fig1}
\end{figure}

Since the discovery of superconductivity at $T_{c}=26$ K in the iron-based
LaO$_{1-x}$F$_{x}$FeAs \cite{26K}, great efforts have been devoted to
explore new kind of iron oxypnictide superconductors with the higher
superconducting transition temperature. As shown in Fig. \ref{fig1}, through
elemental substitution, $T_{c}$ was drastically raised to $43$ K ($>39$ K,
the commonly assumed McMillan Limit) in SmO$_{1-x}$F$_{x}$FeAs \cite{43K}
that can be defined as a unconventional superconductor. One day later,
superconductivity at 41 kelvin in another iron-based layered compound CeO$%
_{1-x}$F$_{x}$FeAs \cite{41K} was reported. Immediately, many group reported
that the $T_{c}$ could be further increased to above 50 K, for example, 52 K
in PrO$_{1-x}$F$_{x}$FeAs \cite{52K}, 55 K in SmO$_{1-x}$F$_{x}$FeAs \cite%
{55K} and 56 K in Gd$_{1-x}$Th$_{x}$FeAsO \cite{56K}. Facing the rapid
increase of $T_{c}$, researchers and the media have become too crazy about
the new discovery \cite{adrian}. This situation is remarkably similar to
that of the discovery of cuprate superconductors in 1986. Some researchers
even claimed that the new discovery may pave the way for the development of
superconductors that can operate at room temperature \cite{room}.

The quest for higher $T_{c}$ iron-based superconductors is still continuing.
For all researchers, it is not known to what extent $T_{c}$ can increase in
these materials. However, on 29 August, we argued that the superconducting
transition temperature of the previously reported series compounds is
difficult to break through $T_{c}=60$ K \cite{huang0}. Later, Steele
concluded that since the metallic bond (conduction band) in the material is
confined to the lines of iron atoms, the $T_{c}$ of any of these materials
will never be as high as in the high-$T_{c}$ cuprates \cite{steele}. Now,
more than two months have passed since our prediction that these materials
could be adjusted by a appropriate charge carrier density to raise theirs $%
T_{c}$ by a few more kelvins. Since then, though several new superconductors
have been discovered \cite{38K,50K,18K,42K,9K,30_1K,36_1K,56_a}, their $%
T_{c} $ are also lower than $60$ K (see Fig. \ref{fig1}) as our suggested.

Some readers are eager to know how can we make such a bold prediction (the $%
T_{c}^{\max }$ of the related materials has to be less than$\ 60$ K)? In the
present paper, on the one hand, we try to answer the question raised above.
On the other hand, the ways how to achieve some higher $T_{c}$ iron-based
superconductors are suggested.

\section{Why we need a fresh thinking to understand superconductivity?}

Superconductivity was used to be considered as a peculiar physical
phenomenon, which can only be observed in few special materials (or
elements). Now, as more and more materials (more than several thousands)
with different structures and physical properties have been discovered to
exhibit superconductivity, the previous viewpoint must be changed. In our
opinion, it is most likely that any crystalloid materials with a appropriate
charge carrier density (not too high, not too low) may found to be the
superconductors under an appropriate temperature. In fact, the field of
superconductivity is in a transition from an old era [Which materials (or
elements) can be a superconductor?] to a new era [Which materials (or
elements) cannot be a superconductor?]. Therefore, it is unnecessary to be
too excited about the observation of the superconductivity in layered iron
arsenic compounds.

In the past century, the experimental scientists have discovered so many
different kinds of superconductors, which greatly challenge the thinking of
the theoretical physicists. What causes the superconductivity? For most
\textquotedblleft theoretical physicists\textquotedblright , it seems
natural that different superconductors will work differently, or we always
need a new mechanism for a new superconductor.\ To speculate the mechanisms
for the various superconductors, they have spend a good deal of time using
computers to compute numerical solutions to equations obtained by the
complicated mathematical derivation. They mistakenly believe that the
reliable results can only be expected by the applying of high-complex
mathematics and advanced computer. We think this approach must be given up.
It is time for us to realize that God didn't create superconductivity in
such a complicated way as people considered to be. We believe that the
superconductivity, as a common phenomenon in nature, its law should be
intrinsically simple and deterministic \cite{unified}. It is physically
unrealistic to expect that the maths and computer can unravel the
superconductivity mystery and tell us the right answer. In fact, the misuse
of the mathematical and computing techniques rather than physics thinking
may obscure the essential physics underlying the superconducting phenomenon.

\begin{figure}[tbp]
\begin{center}
\resizebox{0.85\columnwidth}{!}{
\includegraphics{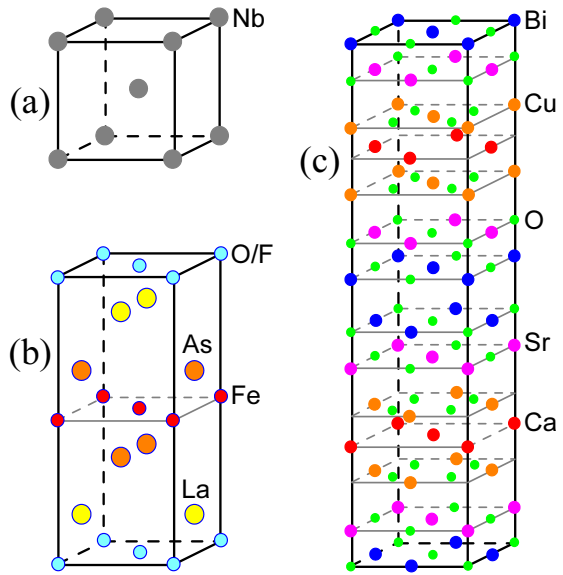}}
\end{center}
\caption{The unit cell for three different superconductors. (a) The
conventional superconductor Nb with a bcc structure, (b) the new iron-based
LaO$_{1-x}$F$_{x}$FeAs, and (c) the cuprate Bi$_{2}$Sr$_{2}$CaCu$_{2}$O$_{8}$%
. }
\label{fig2}
\end{figure}

Figure \ref{fig2} shows the unit cell of three typical superconductors, they
are (a) the conventional superconductor Nb with a bcc structure, (b) the
iron-based LaO$_{1-x}$F$_{x}$FeAs and (c) the cuprate Bi$_{2}$Sr$_{2}$CaCu$%
_{2}$O$_{8}$. For the conventional superconductors, the mainstream physicist
still believe that the BCS theory \cite{BCS} can explain the behavior of
superconductivity. For the cuprate high-$T_{c}$ superconductors \cite%
{bednorz}, a widely-accepted explanation is still missing despite great
effort since 1986. Recently, Anderson \cite{anderson0} even pointed out that
the need for a bosonic glue (phonon) in cuprate superconductors is folklore
rather than the result of scientific logic and many theories about electron
pairing in cuprate superconductors may be on the wrong track. For the new
iron-based superconductors, some researchers believe that the new family
contains new mysteries and some new theoretical models should be invented.

Even though the BCS theory has been proved to be invalid in most of the
known superconductors, this situation has not made researchers rethink one
basic question: Is the BCS theory correct? Most recently, I have discussed
with Prof. Anderson about the reliability of BCS theory \cite{anderson}. He
considered the application of the BCS theory of superconductivity should be
confined within a small domain of materials, the polyelectronic metals. But,
I have showed clearly that the phonon-mediated BCS theory is fundamentally
incorrect (see Section II of Ref.\cite{unified}). I insisted that the BCS
theory is unsuitable not only for the non-conventional superconductors but
also for the conventional superconductors.

It is well known that the BCS theory was established on the basis of quantum
mechanism, which is extreme sensitivity to a small perturbation of the
Hamiltonian of the studied system. One can recall the process of the
establishment of the BCS theory, even for the simplest superconductor of
Fig. \ref{fig2}(a), which included a large number of man-made physical
hypotheses and mathematical approximations (most of which are physically
unreasonable). It is no doubt that, for the complex superconductors of Figs. %
\ref{fig2}(b) and (c), the quantum mechanics is powerless. We have to give
up the original idea and procedure of BCS type, of course, the new theory of
superconductivity should not be too concerned about the specific atomic
structures of the superconductors.

\begin{figure}[bp]
\begin{center}
\resizebox{1\columnwidth}{!}{
\includegraphics{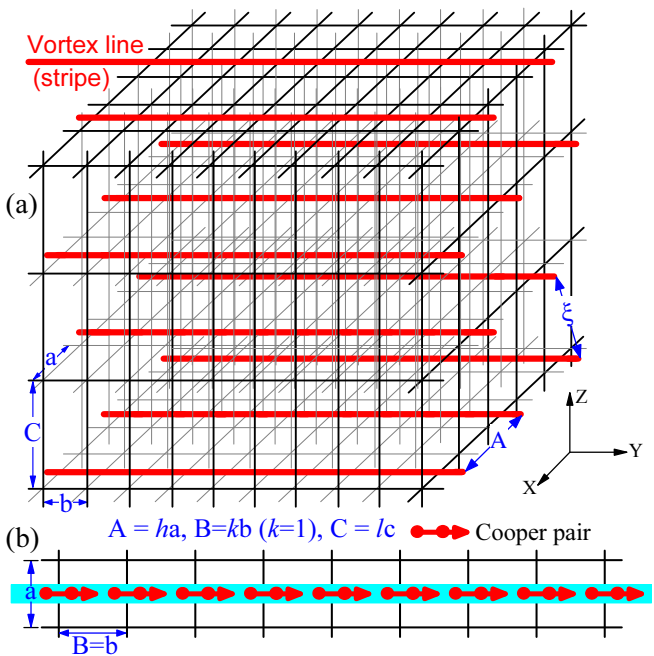}}
\end{center}
\caption{The schematic plot of simplified model of superconductor. (a) The
model includes only two important elements: the framework structure (black
lines) of atoms and the vortex lattice (red lines) of the charge carriers,
(b) the detail structure of vortex line.}
\label{fig3}
\end{figure}

\section{Theory}

Physically, the superconducting state is merely a charge-order phase in the
superconductors. It is unwise to endow superconductivity with too many
mysterious elements. In the framework of BCS theory, the charge carriers are
assumed to be in $k$-space (dynamic screening) order but $r$-space
(real-space screening) disorder. From a strictly mathematical viewpoint, $k$%
-space picture and $r$-space picture are tightly correlated, obviously, the
BCS picture is unwarranted conjecture in both physics and maths. We argue
that a reasonable superconducting theory must be mathematically
self-consistent, in other words, the charge carriers should display a
similar degree of order in both $k$-space and $r$-space. In addition, we
believe that a physical theory in real-space picture is naturally more
reliable than that in momentum-space picture, as all physical phenomena take
place in the real space rather than the imaginary $k$-space.

Recently, we have developed a real space vortex-lattice based theory of
superconductivity which can naturally explain some complicated problems in
conventional and non-conventional superconductors, included the new iron
arsenide superconductors \cite%
{unified,huang0,huang1,huang2,huang3,huang4,huang5}. As shown in Fig. \ref%
{fig3}(a), the mechanism contains two main factors: the framework structure
(black lines) of atoms and the vortex lattice (red lines) of the charge
carriers (electrons). In this case, a real space long range magnetic order
(vortex line, or stripe) and superconductivity coexist to form a dimerized
charge supersolid (a charge-Peierls dimerized transition), as seen in Fig. %
\ref{fig3}(b).
\begin{figure}[tbp]
\begin{center}
\resizebox{1\columnwidth}{!}{
\includegraphics{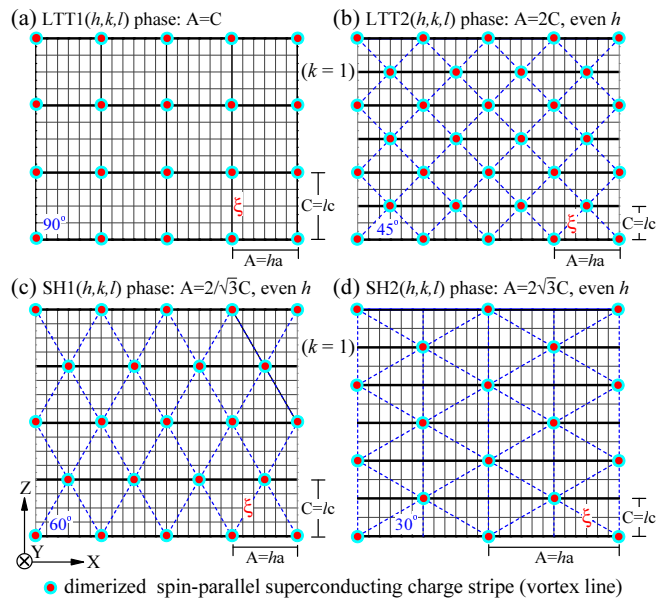}}
\end{center}
\caption{Four possible stable superconducting vortex lattices. (a) A
tetragonal LTT1($h,k,l$) phase, (b) a tetragonal LTT2($h,k,l$) phase, (c) a
trigonal SH1($h,k,l$) phase, and (d) a trigonal SH2($h,k,l$) phase. And the
superlattice constants satisfy: $A=C$, $A=2C$, $A=2/\protect\sqrt{3}C$ and $%
A=2\protect\sqrt{3}C$, respectively.}
\label{fig4}
\end{figure}

In the previous studies\cite{unified,huang0,huang4}, we argued that the
physically significant critical value for the most stable vortex lattice is
that at which a uniform distribution of vortex lines in the plane
perpendicular to the stripes. In this sense, two low-temperature tetragonal
phases (LTT1 and LTT2) and two simple hexagonal phases (SH1 and SH2) might
be the ideal candidates for the stable charge-stripe order of paired
electrons, as shown in Fig. \ref{fig4}.

For a doped superconductor, the charge carrier doping level $x$ is given by

\begin{equation}
x=p(h,k,l)=2\frac{V_{abc}}{V_{ABC}}=2\times \frac{1}{h}\times \frac{1}{k}%
\times \frac{1}{l},
\end{equation}%
and the corresponding charge carrier density is%
\begin{equation}
\rho _{s}=\frac{2}{ABC}=\frac{2}{hkl}\frac{1}{abc}=\frac{x}{abc},
\label{density}
\end{equation}%
where $(A,B,C)=(ha,kb,lc)$, $h$, $k$, and $l$ are integral numbers, and $%
V_{abc}$ and $V_{ABC}$\ are the unit cell volumes of the\ lattice and the
corresponding superlattice, respectively.

It should be pointed out that, for most actual superconducting materials,
their superconducting vortex lattices are likely in some not standard vortex
structures with some degree of distortion compared to the standard
structures of Fig. \ref{fig4}. Obviously, the distortion of the vortex
lattice can affect the stability of the superconducting state, at the same
time decrease the superconducting transition temperature of the
corresponding superconductor. Experimentally, by exerting an external
pressure on the superconductor, or through the substitution of smaller ions
(chemical inner pressure), the vortex lattice distortion can be corrected in
some extent, as a result, increasing the $T_{c}$ of the studied
superconductor. These effects imply that the shrinking of the lattice
constants may enhance the $T_{c}$ of the superconductor.

But in the following discussions we will show that increasing the $c$-axis
lattice constant can result in a much more intensive enhancement of $T_{c}$
in the iron-based superconductors. It must be pointed out that these two
conclusions are not contradictory. Just as in the cuprate superconductors,
although it has been proven that the reducing of the lattice constants\ by
an external pressure through is probable resulting in the increasing of the $%
T_{c}$ in these superconductors, however it is no doubt that the most
effective means of enhancing $T_{c}$ is by means of increasing $c$-axis
lattice constant of the superconductors.\ As we can see that the $T_{c}$ of
the cuprate superconductors had been raise easily from the $T_{c}=40$ K of
the La$_{2-x}$Ba$_{x}$CuO$_{4}$ ($c=13.2\mathring{A}$), to $T_{c}=80$ K of
the YBa$_{2}$Cu$_{4}$O8$_{4}$ ($c=27.24\mathring{A}$), $T_{c}=110$ K of the
Bi$_{2}$Sr$_{2}$Ca$_{2}$Cu$_{3}$O$_{10}$ ($c=37.1\mathring{A}$) and $%
T_{c}=136$ K of the HgBa$_{2}$Ca$_{2}$Cu$_{3}$O ($c=158.3\mathring{A}$). The
potential physical reasons behind these results are still unclear.

According to Fig. \ref{fig3}, the $T_{c}$ of a superconductor is directly
proportional to the stability of the superconducting vortex lattice phase.
There are three factors (temperature, charge carrier density and $c$-axis
lattice constant) that affect vortex lattice's situations, which in turn
influence the $T_{c}$ of a superconductor. Each of these influences will be
discussed in greater detail in the following subsections. \label{temperature}

\begin{figure}[tbp]
\begin{center}
\resizebox{1\columnwidth}{!}{
\includegraphics{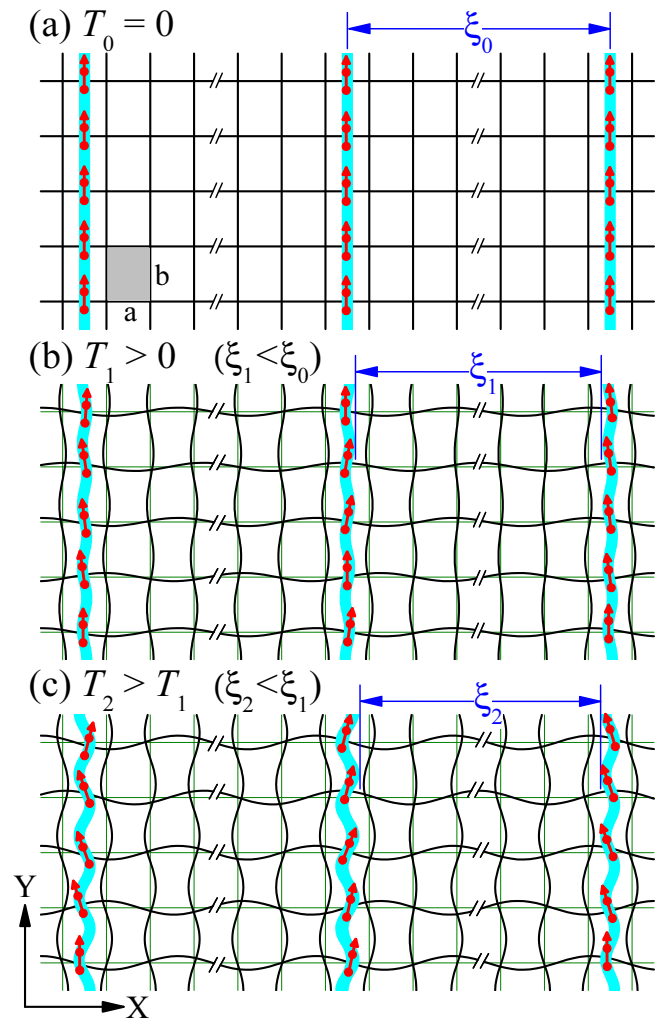}}
\end{center}
\caption{The schematic interpretation of the influence of temperature on the
$T_{c}$ of the superconductors. (a) $T=0$, (b) a lower temperature $T_{1}$,
and (c) a higher temperature $T_{2}>T_{1}$. Obviously, the impact of
temperature on the superconducting transition temperature $T_{c}$ is always
negative.}
\label{fig5}
\end{figure}

\subsection{The influence of temperature on the $T_{c}$}

In the ideal situation of absolute zero temperature, the vibration of the
lattice framework and the flustration of the vortex lines (stripes) can be
neglected, as shown in Fig. \ref{fig5}(a). In this special case, the vortex
lattice is in the most stable minimum energy superconducting state (the
so-called ground state). When $T_{1}>0$, there exist inevitably the
vibration of the lattice framework and the flustration of the vortex lines,
as shown in Fig. \ref{fig5}(b). Furthermore, these may directly lead to a
stronger stripe-stripe interaction due to a shorter minimum stripe-stripe
distance ($\xi _{1}<\xi _{0}$). As illustrated in Fig. \ref{fig5}(c), the
stripe-stripe interaction increased with the increasing of the temperature ($%
T_{2}>T_{1}$). These discussions imply that temperature on the impact of the
superconducting transition temperature $T_{c}$ is always negative and the
lattice vibration (phonon) is impossible to provide the \textquotedblleft
glue\textquotedblright\ for the superconductors \cite{unified}.

\subsection{The influence of charge carrier density on the $T_{c}$}

\label{charge}

\begin{figure}[tbp]
\begin{center}
\resizebox{1\columnwidth}{!}{
\includegraphics{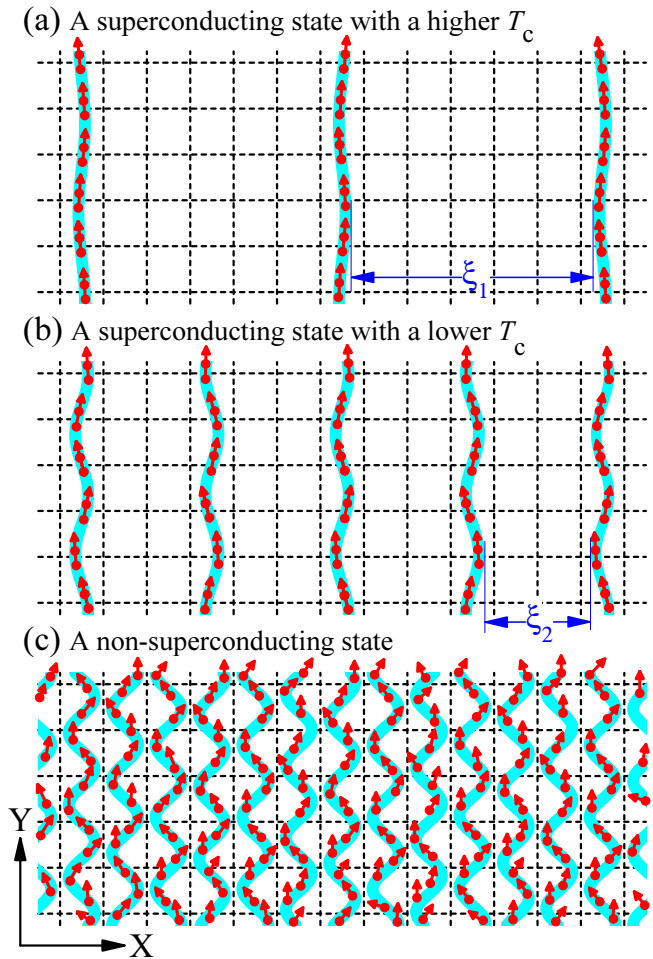}}
\end{center}
\caption{The schematic interpretation of the influence of charge carrier
density on the $T_{c}$ of the superconductors. (a) The superconducting plane
of the low-doped superconductor, (b) a higher doped superconductor, and (c)
a non-superconducting phase due to the excessive charge carrier density.}
\label{fig6}
\end{figure}

It is well known that, in both the cuprate and the new iron-pnictide
high-temperature superconductors, their superconducting transition
temperature $T_{c}$ can be modified by the charge carrier density $\rho _{s}$
(or the doping level $x$). So far, though the influence of the carrier
concentration on the fundamental properties of high-$T_{c}$ superconductors
has been experimentally and theoretically studied by many investigators.
Scientists have not yet reached a consensus on this issue.

In our theoretical framework of Fig. \ref{fig3}, the relationship between
the charge carrier density and $T_{c}$ is very simple and intuitive. As
shown in Fig. \ref{fig6}, in a low doping superconducting vortex phase of
Fig. \ref{fig6}(a), the nearest neighbor distance ($\xi _{1}$) between the
vortex lines is large, so that the stripe-stripe interactions are weak and
the superconductor is expected to have a higher $T_{c}$. When further charge
carriers are added to the superconductor, the vortex lines [see Fig. \ref%
{fig6}(b)] become more crowd than that of Fig. \ref{fig6}(a) ($\xi _{2}<\xi
_{1}$). As a result, a higher charge-carrier density will produce a stronger
interaction among the vortex lines and consequently lead to a lower $T_{c}$.
Figure \ref{fig6}(c) shows an extreme case (with the doping level $x=2$) in
which each unit cell contains two electrons, obviously, the superconducting
vortex lattice no longer exists due to the great enhancement of the
stripe-stripe interactions. These result strongly suggest that a
superconducting phase can be destroyed easily by the extra charge carriers
in a superconductor. We consider the picture of Fig. \ref{fig6}(c) provides
a vivid interpretation: Why the good conductors (Cu, Au, and Ag) and
overdoped high-$T_{c}$ superconductors do not achieve or exhibit
superconductivity? In brief, the superconducting state is characterized by a
real-space periodic vortex lattice while the non-superconducting state is
dominated by the anomalous distribution of charge carriers.

\subsection{The influence of $c$-axis lattice constant on the $T_{c}$}

\label{lattice}

For the cuprate superconductors, there are many experimental facts showing a
strong dependence of the maximum superconducting transition temperature ($%
T_{c}^{\max }$) on the $c$-axis lattice constant. Normally, a superconductor
with a larger $c$-axis lattice constant may have a higher maximum
superconducting transition temperature. But the dependency of $T_{c}^{\max }$
on the $a$-axis (or $b$-axis) lattice constant is much less sensitive
compared to that of the $c$-axis lattice constant.

\begin{figure}[tbp]
\begin{center}
\resizebox{1\columnwidth}{!}{
\includegraphics{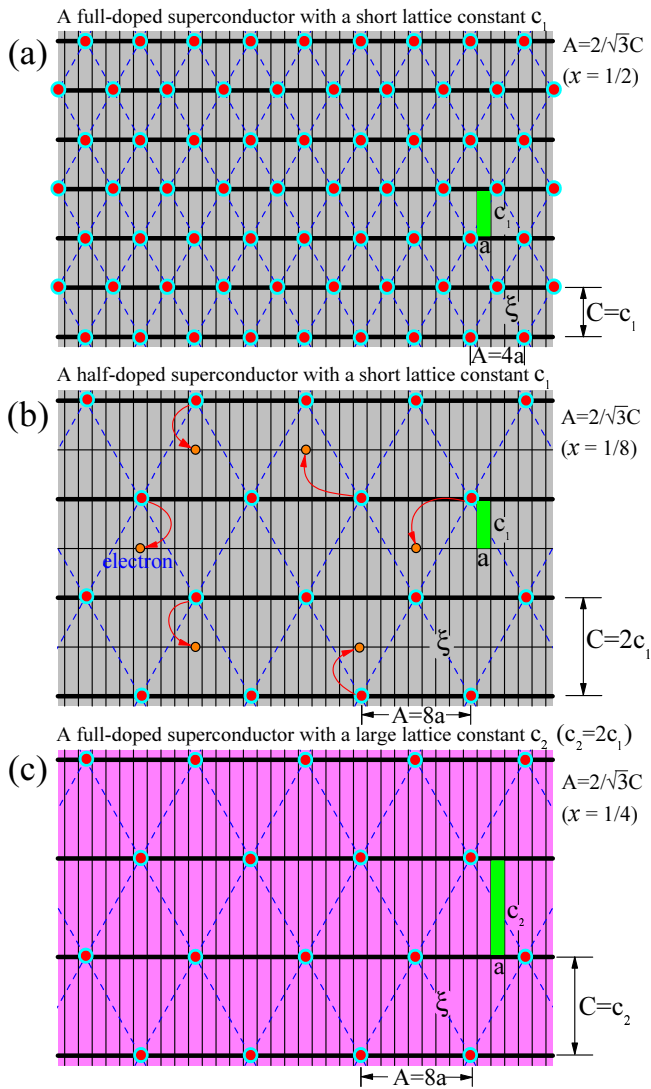}}
\end{center}
\caption{An explanation why a layered material with a large $c$-axis lattice
constant tends to be a high-$T_{c}$ superconductor. (a) A crowded
superconducting vortex lattice in a superconductor with a short $c$-axis
lattice constant, the corresponding $T_{c}$ is low due to a stronger
stripe-stripe interaction among the vortex lines. (b) A relatively uncrowded
vortex lattice in the short $c$-axis superconductor, though the
stripe-stripe interactions could have been reduced greatly, the stability of
the superconducting vortex lattice may be decreased by some mis-site
electrons. (c) A possible higher $T_{c}$ superconducting vortex lattice in a
larger $c$-axis superconductor.}
\label{fig7}
\end{figure}

Here, we try to uncover the possible relationships between the lattice
constant $c$ and $T_{c}^{\max }$ based on the scenario in Fig. \ref{fig4}.
Suppose there are two different kinds of superconductors with the lattice
constants ($a$, $b$, $c_{1}$) and ($a$, $b$, $c_{2}=2c_{1}$) respectively,
as shown in Fig. \ref{fig7}. If $c_{1}=2\sqrt{3}a$, some hexagonal vortex
lattices can be formed in the superconductors at some appropriate doping
levels $x=1/2$, $1/8$ and $1/4$, which are shown in Figs. \ref{fig7}(a), (b)
and (c) respectively. As can be seen from the figures, all the
superconducting layers are doped in the cases of Figs. \ref{fig7}(a) and
(c), it is clear that the former case (for the superconductor having a
shorter $c$-axis parameter) will exhibit a lower $T_{c}$ due to a much
stronger stripe-stripe interaction among the crowded vortex lines. Of
course, by reducing the charge carrier concentration in the superconductor
of Fig. \ref{fig7}(a), it is possible that the two superconductors could
have the same charge carrier density, are shown in Figs. \ref{fig7}(b) and
(c). It should be noted that only half of the superconducting layers are
doped in Fig. \ref{fig7}(b). This inevitably lead to the escapement of
electrons from the vortex lines to the undoped layers, as illustrated in
Fig. \ref{fig7}(b), consequently, decreases the stability of the
superconducting vortex lattice and reduces the corresponding superconducting
transition temperature. For a even shorter $c$-axis lattice constant, for
example $c_{1}^{\prime }=c_{2}/n$ ($n>2$), there will be more undoped layers
in Fig. \ref{fig7}(b) and this may greatly increase the possibility of the
electrons going wrong into these non-superconducting layers, thereby
increase the instability of the superconducting vortex lattice state. In
other words, by separating the planes of the superconducting layers by inert
some non-superconducting layers, the superconducting vortices will be
stabilized to a greater extent , thus raising $T_{c}$ of the superconductor.

It is apparent from the above discussions that a materials with a larger $c$%
-axis lattice constant could potentially be a high-temperature
superconductor. This does not mean that the $T_{c}$ can increase unboundedly
by the increasing of the $c$-axis lattice constant of the superconducting
material. In our mechanism of superconductivity, it is not difficult to find
that a appropriate stripe-stripe interaction is absolutely necessarily for
the formation of the superconducting vortex lattice.

\subsection{The critical energy of the superconductor}

In this subsection, from the viewpoint of energy, we elaborate on the
relationship between the stability of superconducting vortex phase and the
two physical parameters (temperature $T$ and lattice constant $c$).
Physically, the stability of one superconducting phase will decrease with
the increase of the superconductor's energy. In subsection \ref{temperature}%
, it had been shown that the temperature remained unconducive to the
formation of the superconducting vortex state. This conclusion can be
expressed simply by%
\begin{equation}
E_{T}=\alpha T,  \label{energyT}
\end{equation}%
where $E_{T}$ denotes the temperature-dependent energy of the vortex
lattice, $T$ is the temperature and $\alpha $ is a structure and
material-related parameter.

Similarly, the results of the discussions in Subsection\ref{charge} and
Subsection\ref{lattice} can be explicitly presented as follows
\begin{equation}
E_{S}=\beta \frac{1}{c},  \label{energy_c}
\end{equation}%
where $E_{S}$ denotes the (charge carrier density and $c$-axis lattice
constant)-dependent energy of the vortex lattice, $c$ is the $c$-axis
lattice constant and $\beta $ is another material-related parameter.
Combining Eqs. (\ref{energyT}) and (\ref{energy_c}) yields the total energy $%
E(T,c)$ of the vortex lattice%
\begin{equation}
E(T,c)=E_{T}+E_{S}=\alpha T+\beta \frac{1}{c}.  \label{total}
\end{equation}

Let $T_{c}$ be the superconducting transition temperature, then we can
define the critical energy of the superconductor\ as%
\begin{equation}
E_{C}=E(T_{c},c)=\alpha T_{c}+\beta \frac{1}{c},  \label{critical}
\end{equation}%
here we assume that the $E_{c}$ value is same for any superconductors. And
the superconducting vortex phase should satisfy the following criterion%
\begin{equation*}
E(T,c)-E(T_{c},c)\leq 0\text{,}
\end{equation*}%
while for the non-superconducting phase, we have
\begin{equation*}
E(T,c)-E(T_{c},c)>0.
\end{equation*}

In the following sections, we will apply the above general analysis to the
specific iron-based superconductors.
\begin{table}[tbp]
\caption{Experimental data of $T_{c}$, lattice constants ($a$ and $c$) and $%
c/a$ ratio for the iron-based superconductors with $c/a\approx 3$, where $a $
($=a_{0}/\protect\sqrt{2}$) is the nearest-neighbor $Fe$-$Fe$ distances.}
\label{table1}%
\begin{tabular}{c|c|c|c|c}
\hline\hline
Superconductors & $a($\AA $)$ & $c($\AA $)$ & $c/a$ & $T_{c}(K)$ \\
\hline\hline
\cite{26K}LaO$_{1-x}$F$_{x}$FeAs & $2.850$ & $8.739$ & $3.066$ & $26$ \\
\hline
\cite{43K}SmO$_{1-x}$F$_{x}$FeAs ($x=0.15$) & $2.786$ & $8.496$ & $3.050$ & $%
43$ \\ \hline
\cite{41K}CeO$_{1-x}$F$_{x}$FeAs ($x=0.16$) & $2.820$ & $8.631$ & $3.061$ & $%
41$ \\ \hline
\cite{52K}PrO$_{1-x}$F$_{x}$FeAs ($x=0.11$) & $2.818$ & $8.595$ & $3.050$ & $%
52$ \\ \hline
\cite{55K}SmO$_{1-x}$F$_{x}$FeAs ($x=0.1)$ & $2.768$ & $8.428$ & $3.045$ & $%
55$ \\ \hline
\cite{56K}Gd$_{1-x}$Th$_{x}$FeAsO ($x=0.2$) & $2.769$ & $8.447$ & $3.051$ & $%
56$ \\ \hline
\cite{50K}Tb$_{1-x}$Th$_{x}$FeAsO ($x=0.2$) & $2.759$ & $8.412$ & $3.049$ & $%
50$ \\ \hline
\cite{42K}TbFeAsO$_{0.85}$ & $2.750$ & $8.376$ & $3.046$ & $42$ \\ \hline
\cite{52K_a}SmO$_{1-x}$F$_{x}$FeAs ($x=0.35$) & $2.778$ & $8.522$ & $3.068$
& $52$ \\ \hline
\cite{36_6K}GdO$_{1-x}$F$_{x}$FeAs ($x=0.17$) & $2.829$ & $8.650$ & $3.058$
& $36.6$ \\ \hline
\cite{54_6K}SmO$_{1-x}$F$_{x}$FeAs ($x=0.3$) & $2.777$ & $8.482$ & $3.054$ &
$54.6$ \\ \hline
\cite{54K}SmO$_{1-x}$F$_{x}$FeAs ($x=0.2$) & $2.775$ & $8.481$ & $3.056$ & $%
54$ \\ \hline
\cite{30_1K}LaO$_{0.8}$F$_{0.2}$FeAs$_{1-x}$Sb$_{x}$ ($0.05)$ & $2.843$ & $%
8.701$ & $3.061$ & $30.1$ \\
LaO$_{0.8}$F$_{0.2}$FeAs$_{1-x}$Sb$_{x}$ $(0.10)$ & $2.845$ & $8.719$ & $%
3.065$ & $28.6$ \\ \hline
\cite{zaren}SmFeAsO$_{1-x}$ ($x=0.15)$ & $2.756$ & $8.407$ & $3.050$ & $55$
\\
GdFeAsO$_{1-x}$ ($x=0.15$) & $2.760$ & $8.453$ & $3.063$ & $53.5$ \\
NdFeAsO$_{1-x}$ ($x=0.15$) & $2.788$ & $8.521$ & $3.056$ & $53.5$ \\
PrFeAsO$_{1-x}$ ($x=0.15$) & $2.806$ & $8.566$ & $3.053$ & $51.3$ \\
CeFeAsO$_{1-x}$ ($x=0.15$) & $2.814$ & $8.605$ & $3.058$ & $46.5$ \\
LaFeAsO$_{1-x}$ ($x=0.15$) & $2.844$ & $8.707$ & $3.062$ & $31.2$ \\ \hline
\cite{36_1K}Sr$_{1-x}$La$_{x}$FeAsF & $2.826$ & $8.961$ & $3.171$ & $36.1$
\\ \hline
\cite{56_a}Sr$_{1-x}$Sm$_{x}$FeAsF ($x=0.5$) & $2.825$ & $8.961$ & $3.172$ &
$56$ \\ \hline\hline
\end{tabular}%
\end{table}

\section{The maximum $T_{c}$ in the iron-based superconductors}

Up to now, there are about dozens of iron-based superconductors have been
reported. Based on our mechanism of superconductivity, these superconductors
can be divided into two basic classes ($c/a\approx 3$ and $c/a\approx
5/2=2.5 $) according to the $c$-axis lattice constants. It will be shown
that the $c/a\approx 3$ class has a higher $T_{c}^{\max }$ than that of $%
c/a\approx 5/2 $ class due to a larger $c$-axis lattice constant in the
former class.

Table \ref{table1} shows the experimental data of $T_{c}$, lattice constants
($a$ and $c$) and $c/a$ ratio of the majority reported iron-based
superconductors. It should be noted that in this paper the lattice constant $%
a$ ($=a_{0}/\sqrt{2}$) is the Fe-Fe distance in the $xy$-plane (the
superconducting plane). It is easy to find that, except the last two
samples, all the compounds have a similarity $c/a$ value of 3. According to
the value of $c$-axis lattice constants, the other reported iron-based
superconductors can be classify into another class, as shown in Table \ref%
{table2}. In the previous paper \cite{huang5}, we have mentioned that the $c$%
-axis lattice constant of the Fe$_{2}$As$_{2}$ family should be redefined as
$c=c_{0}/2$, where $c_{0}$ is the corresponding experimental value of the $c$%
-axis lattice constant. It is apparent from this table that all the $c/a$
values are very close to 2.5.

\begin{table}[tp]
\caption{Experimental data of $T_{c}$, lattice constants ($a$ and $c$) and $%
c/a$ ratio for the iron-based superconductors of $c/a\approx 5/2$. Note that
for the Fe$_{2}$As$_{2}$ family, the $c$-axis lattice constant is given by $%
c=c_{0}/2$, where $c_{0}$ is the experimental value. }
\label{table2}%
\begin{tabular}{c|c|c|c|c}
\hline\hline
Superconductors & $a($\AA $)$ & $c($\AA $)$ & $c/a$ & $T_{c}(K)$ \\
\hline\hline
\cite{37_2}Sr$_{1-x}$Cs$_{x}$Fe$_{2}$As$_{2}$ ($x=0.4$) & $2.765$ & $6.880$
& $2.488$ & $37.2$ \\
($x=0.5$) & $2.765$ & $6.883$ & $2.489$ &  \\ \hline
\cite{37_2}Sr$_{1-x}$K$_{x}$Fe$_{2}$As$_{2}$ ($x=0.4$) & $2.751$ & $6.474$ &
$2.353$ & $36.5$ \\
($x=0.5$) & $2.731$ & $6.777$ & $2.482$ &  \\ \hline
\cite{38K}Ba$_{1-x}$K$_{x}$Fe$_{2}$As$_{2}$ ($x=0.4$) & $2.764$ & $6.606$ & $%
2.391$ & $38$ \\
$x=1$ &  &  &  & $3$ \\ \hline
\cite{32K}Eu$_{1-x}$K$_{x}$Fe$_{2}$As$_{2}$ ($x=0.5$) & $2.734$ & $6.5455$ &
$2.394$ & $32$ \\ \hline
\cite{18K}LiFeAs & $2.681$ & $6.364$ & $2.374$ & $18$ \\
& $2.669$ & $6.354$ & $2.381$ & $16$ \\ \hline
\cite{9K}NaFeAs & $2.791$ & $6.9911$ & $2.505$ & $9$ \\ \hline\hline
\end{tabular}%
\end{table}

\begin{figure}[tbp]
\begin{center}
\resizebox{1\columnwidth}{!}{
\includegraphics{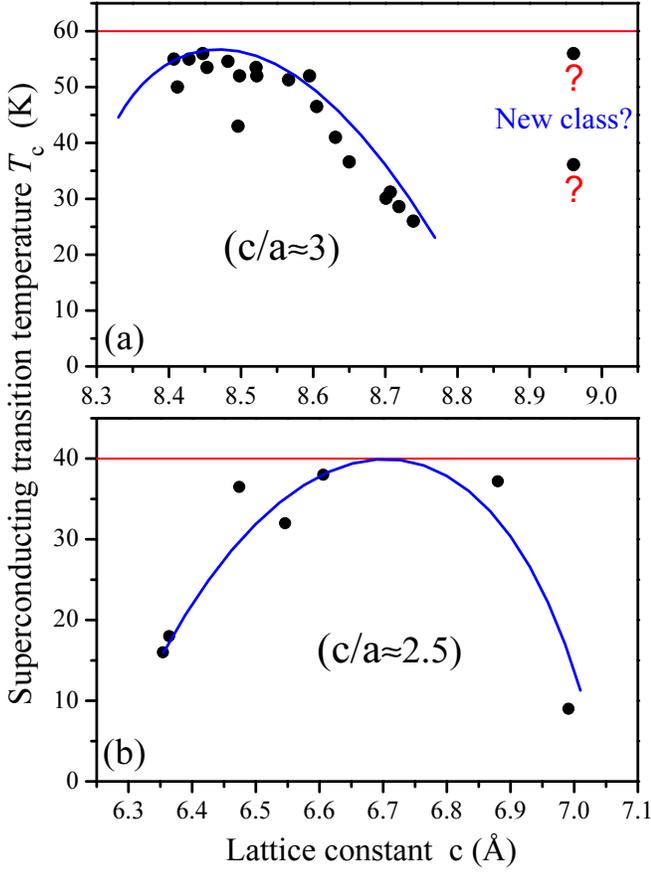}}
\end{center}
\caption{The relationship between $T_{c}$ and the $c$-axis lattice constants
based on the reported experimental data. (a) For the large $c$ iron-based
superconductors, and (b) for the short $c$ iron-based superconductors. Our
predictions of the maximum $T_{c}$ for both classes are also shown in the
figures.}
\label{fig8}
\end{figure}

The relationships between the $c$-axis lattice constants and $T_{c}$\ are
depicted in Fig. \ref{fig8} for both classes. Let us look at Fig. \ref{fig8}%
(a) of $c/a\approx 3$ class, the blue curve indicates that the maximum $%
T_{c}^{\max }$ of this group is difficult to exceed 60 K. In addition, there
are two exceptional values, as marked in the subfigure, which are most
likely to belong to a new class of $c/a\approx 16/5$. For the $c/a\approx
5/2 $ class, as shown in Fig. \ref{fig8}(b), the highest superconducting
transition temperature is about 40 K due to the relatively small $c$-axis
lattice constants.

\section{How to achieve a higher $T_{c}$ in the iron-based superconductors?}

By adjusting the sample carrier concentration and improving the sample
quality, it may still be possible to raise the $T_{c}$ of the iron-based
family by a few more kelvins. According to our theory, increasing the $c$%
-axis lattice constant appeared to be the most effective method of enhancing
$T_{c}$ in FeAs family. In what follows, we will estimate roughly the $c$%
-axis lattice constant values that can make the $T_{c}$ of iron-based
materials into the liquid nitrogen range ($77$ K), and even $100$ K.
\begin{figure}[tbp]
\begin{center}
\resizebox{0.98\columnwidth}{!}{
\includegraphics{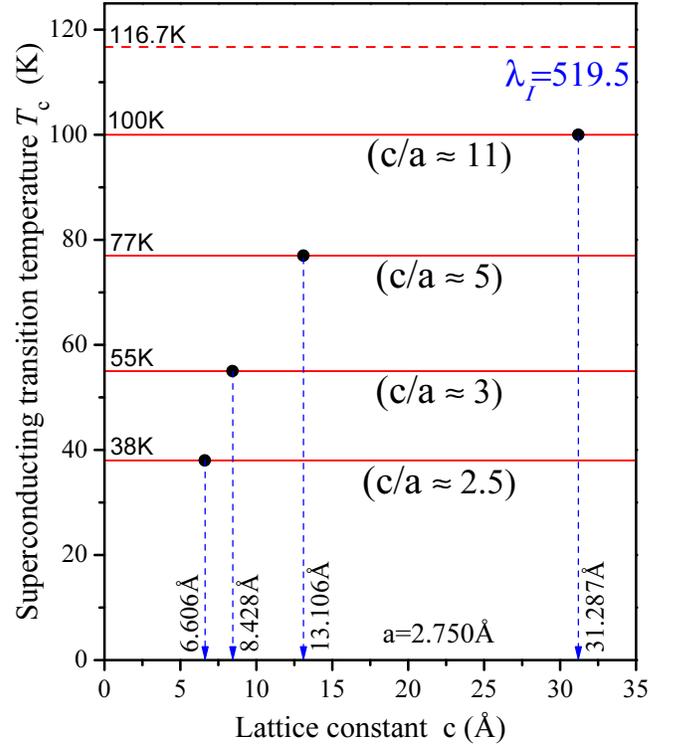}}
\end{center}
\caption{The theoretical results of the relationship between the $%
T_{c}^{\max }$ and the lattice constants for the iron-based superconductors.
The $c$-axis lattice constant plays a key role in promoting the
superconducting transition temperature. }
\label{fig9}
\end{figure}

According to the lattice constants, we now have two classes of the
iron-based superconductors: the $c/a\approx 3$ class and the $c/a\approx 5/2$
class. From the experimental data of Table \ref{table1} and Table \ref%
{table2}, we can obtain two characteristic parameters for each class. They
are $T_{c}^{\max }(1)=55$ K and $c_{1}=8.428\mathring{A}$ for the $%
c/a\approx 3$ class and $T_{c}^{\max }(2)=38$ K and $c_{2}=6.606\mathring{A}$
for the $c/a\approx 5/2$ class. By applying the hypothesis of Eq. (\ref%
{critical}), we have

\begin{eqnarray}
E\left[ T_{c}(1),c_{1}\right] &=&E\left[ T_{c}(2),c_{2}\right] ,  \notag \\
\alpha _{1}T_{c}^{\max }(1)+\beta _{1}\frac{1}{c_{1}} &=&\alpha
_{2}T_{c}^{\max }(2)+\beta _{2}\frac{1}{c_{2}},  \label{ab}
\end{eqnarray}%
where ($\alpha _{1}$,$\beta _{1}$) and ($\alpha _{2}$,$\beta _{2}$) are
material-related parameters.

For ease of discussion, we further assume $\alpha _{1}=\alpha _{2}=\alpha
_{I}$ and $\beta _{1}=\beta _{2}=\beta _{I}$, then we can redefine a unified
parameter by the using of Eq. (\ref{ab})
\begin{eqnarray}
\lambda _{I} &=&\frac{\beta _{I}}{\alpha _{I}}=\frac{c_{1}c_{2}}{c_{2}-c_{1}}%
\Delta T_{c}^{\max }  \label{parameter} \\
&=&\frac{c_{1}c_{2}}{c_{2}-c_{1}}\left[ T_{c}^{\max }(2)-T_{c}^{\max }(1)%
\right] .  \notag
\end{eqnarray}%
By applying the experimental data, one can obtain
\begin{equation*}
\lambda _{I}=\frac{8.428\times 6.606\times (55-38)}{(8.428-6.606)}=519.5.
\end{equation*}

We consider that the estimated value of $\lambda _{I}=519.5$ is the general
characteristic parameter of the iron-based superconductors. This parameter
may help point the way to finding variants of the FeAs materials with higher
$T_{c}$. Eq. (\ref{parameter}) can be rewritten as
\begin{equation*}
c_{2}=\frac{\lambda _{I}\times c_{1}}{\lambda _{I}-\left[ T_{c}^{\max
}(2)-T_{c}^{\max }(1)\right] \times c_{1}}.
\end{equation*}

From the above equation, we now have three known parameters: $c_{1}=8.428$ $%
\mathring{A}$, $T_{c}^{\max }(1)=55$ K and$\ \lambda _{I}=519.5$. For a
given target superconducting transition temperature$T_{c}^{\max }(2)$, it is
very easy to get the corresponding $c$-axis lattice constant $c_{2}$. Figure %
\ref{fig9} shows the results for two target temperatures:$\ T_{c}^{\max
}(2)=77$ K and $100$ K. These results indicate that in order to raise $T_{c}$
into the liquid nitrogen range ($77$ K), the $c$-axis lattice constant
should at least reach $13$ $\mathring{A}$. For even higher $T_{c}$, for
example $T_{c}=100$ K, the $c$-axis lattice constant is estimated to be
around $31.287$ $\mathring{A}$. But in any case, the maximum superconducting
transition temperature ($T_{c}^{\max }$) of the iron-based superconductors
would not exceed 116.7 K, as indicated in Fig. \ref{fig9}.

\section{The latest experimental and theoretical results}

Recently, the new type of Fe superconductor with a $T_{c}$ of approximately
8 K has been discovered for tetragonal FeSe compound \cite{8K}. Later the $%
T_{c}$ of 27 K for the FeSe superconductor at 1.48 GPa, showing an extremely
high pressure coefficient of 9.1 K/GPa, has been reported \cite{27K}. The
crystal structure of the FeSe is the simplest among the reported iron-based
superconductors with the shortest $c$-axis lattice constant $c\approx 5.52$ $%
\mathring{A}$\cite{27K} and the smallest value of $c/a\approx 2$, where $%
a=3.7696/\sqrt{2}\mathring{A}$. From our mechanism, the FeSe-related
superconductors has the lowest $T_{c}^{\max }$ due to a relatively shorter $%
c $-axis lattice constant. The corresponding $T_{c}^{\max }(2)$ can be
estimated by
\begin{equation*}
T_{c}^{\max }(2)=T_{c}^{\max }(1)+\frac{\lambda _{I}\times (c_{2}-c_{1})}{%
c_{1}\times c_{2}}.
\end{equation*}

Roughly, we use $T_{c}^{\max }(1)=55$ K, $c_{1}=8.428\mathring{A}$, $%
c_{2}\approx 5.52$ $\mathring{A}$ and $\lambda _{I}=519.5$ and obtain $%
T_{c}^{\max }(2)\approx 24$ K, which is very close to the experiment value
of 27 K\cite{27K}. Even taking into account all possible factors, we
consider that the $T_{c}^{\max }$ of the FeSe family is impossible to exceed
$30$ K because of the limitation of the short $c$-axis lattice constant.
\begin{figure}[tbp]
\begin{center}
\resizebox{1\columnwidth}{!}{
\includegraphics{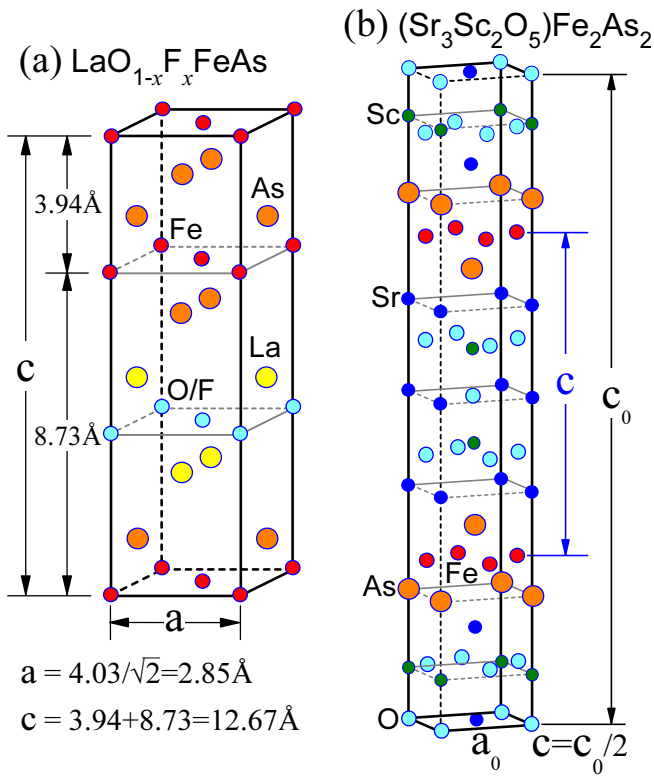}}
\end{center}
\caption{The lager $c$-axis unit cell discovered in (a) the LaO$_{1-x}$F$%
_{x} $FeAs superconductor, and (b) the (Sr$_{3}$Sc$_{2}$O$_{5}$)Fe$_{2}$As$%
_{2}$ parent compound for the FeAs-based superconductors.}
\label{fig10}
\end{figure}

We have been concerned about two interesting papers published on the
ArXiv.org\textit{\ }\cite{long_c,zhu}. In the LaO$_{1-x}$F$_{x}$FeAs
Superconductor \cite{long_c}, Shekhar et al. found that apart from the
standard phase ($c=8.716$ $\mathring{A}$) there exists another structural
phase with a larger $c$-axis lattice constant of $c=12.67$ $\mathring{A}$ ($%
c/a\approx 4.5)$ as shown in Fig. \ref{fig10} (a). If the authors can
successfully synthesize some pure samples of $c=12.67$ $\mathring{A}$ around
the following doping levels $x=1/8=0.125$, $2/15\approx 0.133$, $2/9\approx
0.222$ and $2/5=0.4$, we are confident that their $T_{c}$ can easily break
through 60 K, perhaps reach about $T_{c}=70$ K. At the same time, Zhu et al.
fabricated a possible new parent compound for the FeAs-based superconductors%
\cite{zhu}, namely (Sr$_{3}$Sc$_{2}$O$_{5}$)Fe$_{2}$As$_{2}$ with the
lattice constants $a=2.876\mathring{A}$ ($4.0678/\sqrt{2}$) and $c=13.424%
\mathring{A}$ ($26.8473/2$) ($c/a\approx 4.67$), as shown in Fig. \ref{fig10}
(b). Based on the new concept of superconductivity, we suggest the synthesis
of some new related compounds (Re$_{x}$Sr$_{3-x}$Sc$_{2}$O$_{5}$)Fe$_{2}$As$%
_{2}$ (Re = La, Sm, Gd, Ce, Nd, $\ldots $) at the following values of $%
x=2/3\approx 0.667$, $2/5=0.4$ and $1/4=0.25$. The superconducting
transition temperature of these samples is likely close to or even exceed $%
T_{c}=70$ K.

Interestingly, from a chemist's viewpoint, Steele \cite{steele1,steele2}
considered that the iron-based materials\ (PbO)$_{x}$FeAs, (BaO)$_{x}$FeAs
and (BaO)$_{x}$FeP ($x=1,2$ or 3) materials may also be the candidate of the
higher $T_{c}$ superconductors.

\section{Brief summary}

We have studied the maximum superconducting transition temperature ($%
T_{c}^{\max }$) problems in the newly discovered iron-based superconductors.
It has been shown clearly that all the reported FeAs superconductors can be
divided into two basic classes depending on the lattice constants, they are
the $c/a\approx 3$ class of the larger $c$-axis lattice constant and the $%
c/a\approx 5/2$ class of the shorter $c$-axis lattice constant. The results
indicated that the former class has a maximum $T_{c}^{\max }<60$ K, while
the latter class has a lower $T_{c}^{\max }\leq 40$ K. In order to enhance $%
T_{c}^{\max }$ in the iron-based superconductors, new class of FeAs
compounds with a larger ratio of $c/a$ and $c$-axis lattice constant are
suggested. We pointed out that their $T_{c}^{\max }$ values could be raised
into the liquid nitrogen range ($77$ K) and $100$ K, supposing the new
analogues with $c/a\approx 5$ (approximately $c>13$ $\mathring{A}$, if $%
a=2.750$ $\mathring{A}$) and $c/a\approx 11$ ($c>31$ $\mathring{A}$) can be
successfully synthesized in laboratory. Finally, the possible new ways to
raise $T_{c}$ in the iron-based superconductor have been proposed based on
the new experimental results.

\section{Acknowledgments}

The author would like to thank Philip W. Anderson and Robert B. Steele for
many useful discussions.

\end{document}